\documentclass[twocolumn,showpacs,preprintnumbers,amsmath,amssymb]{revtex4}

\usepackage{graphicx}% Include figure files
\usepackage{dcolumn}% Align table columns on decimal point
\usepackage{bm}% bold math

\begin{document}

\preprint{APS/123-QED}

\title{Electron cyclotron maser emission mode coupling to the z-mode on a longitudinal density gradient in the context of solar type III bursts}% Force line breaks with \\

\author{R. Pechhacker}
 %\email{r.pechhacker@qmul.ac.uk}%Lines break automatically or can be forced with \\
\author{D. Tsiklauri}%
 %\email{d.tsiklauri@qmul.ac.uk}
\affiliation{%
School of Physics and Astronomy, Queen Mary University of London, London E1 4NS, United Kingdom
}%

\date{\today}% It is always \today, today,

\begin{abstract}
A beam of super-thermal, hot electrons was injected into maxwellian plasma with a density gradient along a magnetic field line. 1.5D particle-in-cell simulations were carried out which established that the EM emission is produced by the perpendicular component of the beam injection momentum. The beam has a positive slope in the distribution function in perpendicular momentum phase space, which is the characteristic feature of a cyclotron maser. The cyclotron maser in the overdense plasma generates emission at the electron cyclotron frequency. The frequencies of generated waves were too low to propagate away from the injection region, hence the wavelet transform shows a pulsating wave generation and decay process. The intensity pulsation frequency is twice the relativistic cyclotron frequency. Eventually, a stable wave packet formed and could mode couple on the density gradient to reach frequencies of the order of the plasma frequency, that allowed for propagation. The emitted wave is likely to be a z-mode wave. The total electromagnetic energy generated is of the order of $0.1$\% of the initial beam kinetic energy. The proposed mechanism is of relevance to solar type III radio bursts, as well as other situations, when the injected electron beam has a non-zero perpendicular momentum, e.g. magnetron.
\end{abstract}

\maketitle

%\section{\label{sec:intro}Introduction}
Electron beam injection is common in many plasma situations. Examples are solar flares \cite{2004JGRA..10910104E}, cavity magnetrons that generate microwaves \cite{2005TePhL..31..388V,1976PhRvL..37..379B} or various radar applications. Beams of relativistic, hot electrons are also likely to be the source of solar type III radio bursts. In the common theory their parallel momentum is causing Langmuir wave growth through the 'bump-on-tail' instability, and subsequent electromagnetic (EM) wave generation (and emission) via three main possibilities 1) non-linear wave-wave interactions \cite{1958SvA.....2..653G}, 2) linear mode conversion on density gradients \cite{2008PhPl...15j2110K}, and 3) the antenna mechanism \cite{2010JGRA..11501101M}. Recently, particle-in-cell simulations have been carried out on this topic \cite{2005ApJ...622L.157S,2011PhPl...18e2903T}. The above mentioned mechanisms contain contribution of the parallel component of the beam momentum, however, in situations, where the beam is not completely aligned with the magnetic field, there is likely to be a contribution from the perpendicular component. It is conceivable that non-field aligned, unstable electron beam distributions are possible in the solar atmosphere under some conditions. In this case, a bump of the distribution function will form in the perpendicular direction of momentum space. If the bump is strong enough to achieve a positive slope, the cyclotron maser is triggered. The cyclotron maser is known to generate EM emission in x-,o-, and z-mode, depending on the ratio of plasma frequency to electron gyrofrequency \cite{1986ApJ...307..808W}. It has been studied by PIC simulations in the context of the auroral kilometric radiation in Ref.\cite{1999JGR...10410317P}.\\
A beam of fast electrons is injected perpendicularly to the magnetic field into a 1.5D plasma, in order to turn off all contributions from Langmuir waves and analyse the purely EM signal. Time-distance plots as well as time evolution of emission wavelet transforms and distribution function is investigated. A 1.5D maxwellian plasma is considered, allowing spatial variation in $x$ only, while EM fields and particle momenta have all three components. In the solar atmosphere, typical values for the important ratio $\frac{\omega_{ce}}{\omega_{pe}}$ are $0.1 - 10^{-3}$. We choose our parameters as to satisfy this requirement. The background magnetic field is constant $B=B_x=0.0003$T$=3$G, setting the electron gyrofrequency to $\omega_{ce}=5.28 \times 10^7$Hz rad everywhere. We ignore the radial decrease of the magnetic field in first approximation, while keeping a density gradient, which we believe to be crucial for the mode coupling mechanism here. Moreover, the constraint of $\nabla \cdot {\bf B} =0$ in 1.5D does not allow for $B=B(x)$ given that $y-$ and $z-$ components are ignorable. The background temperature is $T=3 \times 10^5$K and isotropic. The background plasma density at $x=0$ is $n_0=10^{14}$m$^{-3}$, giving $\omega_{pe}=5.64 \times 10^8$Hz rad. This sets $\frac{\omega_{ce}}{\omega_{pe}}=0.0935 \ll 1$. Further, the electron Debye length is $\lambda_{De}=3.78 \times 10^{-3}$m, while the grid size is $\lambda_{De}/2$. The simulation setup is such that we study a single magnetic field line connecting Sun and earth. It is predicted that in the limit $r \gg R_{\bigodot}$ the plasma density $n_e(r) \propto r^{-2}$ \cite{1999A&A...348..614M}, therefore, we use the following density profile
%----------------------------------------------------------------------------
\begin{equation}\label{eq:normaln}
  n_{e,i}(x) = n_0 \left[ \left( \frac{x-x_{max}/2}{x_{max}/2 + n_+} \right)^2 + n_- \right]%
\end{equation}
%----------------------------------------------------------------------------
with $x_{max}$ being the total system length, mimicking the distance of the Sun to the earth, and
%----------------------------------------------------------------------------
\begin{eqnarray}
  n_+ = \frac{x_{max}}{2} \frac{1-\sqrt{1-n_-}}{\sqrt{1-n_-}}, & n_- = 10^{-8}. %
\end{eqnarray}
%----------------------------------------------------------------------------
In order to use periodic boundary conditions, the region $x_{max}/2 \leq x \leq x_{max}$ is set to an appropriate density increase. Data analysis will, therefore, focus on the region $0 \leq x \leq x_{max}/2$ only. The simulation is carried out using EPOCH, a fully electromagnetic, relativistic PIC code.\\
A beam of fast electrons is injected perpendicularly to the magnetic field at simulation time $t=0$. This is because Ref.\cite{PoP} established that the parallel momentum (and the associated Langmuir waves) plays no role in the generation of EM emission. The beam carries a total momentum of $p_b=p_{y}=m_e \gamma \frac{c}{2}$ with $\gamma \approx 1.155$. The beam temperature is $T_b=6 \times 10^6$K. The peak beam density is $n_{b0}=10^{11}$m$^{-3}$, while its spatial profile is defined as $n_b(x) = n_{b0} \exp(-[(x-x_{max}/25)/(x_{max}/40)]^8)$
%----------------------------------------------------------------------------
\begin{figure}[htbp]
\includegraphics[scale=0.49]{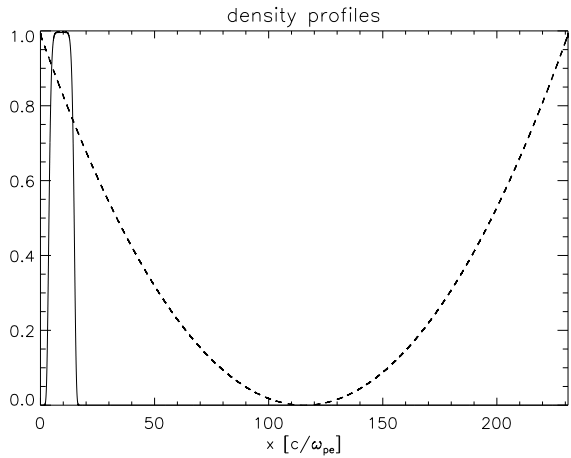}%
\caption{\label{fig:densprofs} beam (solid) and background (dashed) density profiles at $t=0$. Densities are normalized to their maxima.}
\end{figure}
%----------------------------------------------------------------------------
with the beam density maximum at $x_{max}/25$. Fig.\ref{fig:densprofs} shows the density profiles for $t=0$. Note that the beam is injected only at $t=0$. In total we simulate up to $t=150\omega_{pe}^{-1}$. As a result of the above defined quantities, at $x=0$, the plasma beta is $\beta=0.0115$. The mass ratio is $m_i/m_e=1836$. The distribution functions are initially maxwellian. With $p_{\parallel} = p_x$, $p_{\perp} = \sqrt{p_y^2 + p_z^2}$, this leads to $f_{b 0}({\bf p}) = \tilde{n}_{b} \exp(-[p_{\parallel}^2+(p_{\perp}-p_{b\perp})^2]/(2 m_{e} k_B T_b))$ for the beam, where $k_B$ is Boltzmann's constant and $\tilde{n}_b$ is a normalization constant.\\
For a $\theta=90^{\circ}$ beam injection angle with respect to the magnetic field, in a spatial 1D simulation, the beam is trapped at its injection point. Fig.\ref{fig:tdds90} shows time-distance plots for EM field components, $E_x$ and $E_y$, and changes in plasma density (left column), as well as spatial wavelet transforms of snapshots of $E_y$ at times $t=1.8,80,150\omega_{pe}^{-1}$ (right column), referring to the first pulsation maximum (top), the point where the wave packet passes the local plasma frequency (middle), and free propagation (bottom). Electric field strengths are given in units of $\omega_{pe}cm_e/e$, while distance and time are measured in $c/\omega_{pe}$ and $\omega^{-1}_{pe}$, accordingly, where $\omega_{pe}$ is the plasma frequency at $x=0$. The wavelet software was provided by C. Torrence and G. Compo \cite{22}. The wavelet transforms can be interpreted in the same way as described in section IV of Ref.\cite{PoP} and in more detail in Ref.\cite{2011PhPl...18e2903T}\\
Conventional plasma emission processes rely on the generation of Langmuir waves in order to invoke EM emission. Such a signal would be found in the parallel component of the electric field $E_x$, as shown in Ref.\cite{2011PhPl...18e2903T}. It is evident from Fig.\ref{fig:tdds90}, that there is no electrostatic Langmuir signal produced, yet clear wave structures can be found in the transverse components of the EM field, see $E_y$ panel. The wave front of those transverse signals suggest propagation of a signal with the speed of light (units are chosen such that a slope of $\approx 1$ corresponds to a propagation at the speed of light), therefore, the transverse emission is electromagnetic. The lack of the Langmuir signal excludes the possibility that the generation of the EM emission stems from 'conventional' plasma emission mechanisms as mentioned in the introduction. This is due to the fact that in 1.5D {\it plasma emission} is not possible \cite{2011PhPl...18e2903T}. Contribution from the antenna mechanism is also excluded by the fact that there is no density cavity created at any time in the simulation, as shown in the bottom left panel of Fig.\ref{fig:tdds90}.
%----------------------------------------------------------------------------
\begin{figure*}[htbp]
\includegraphics{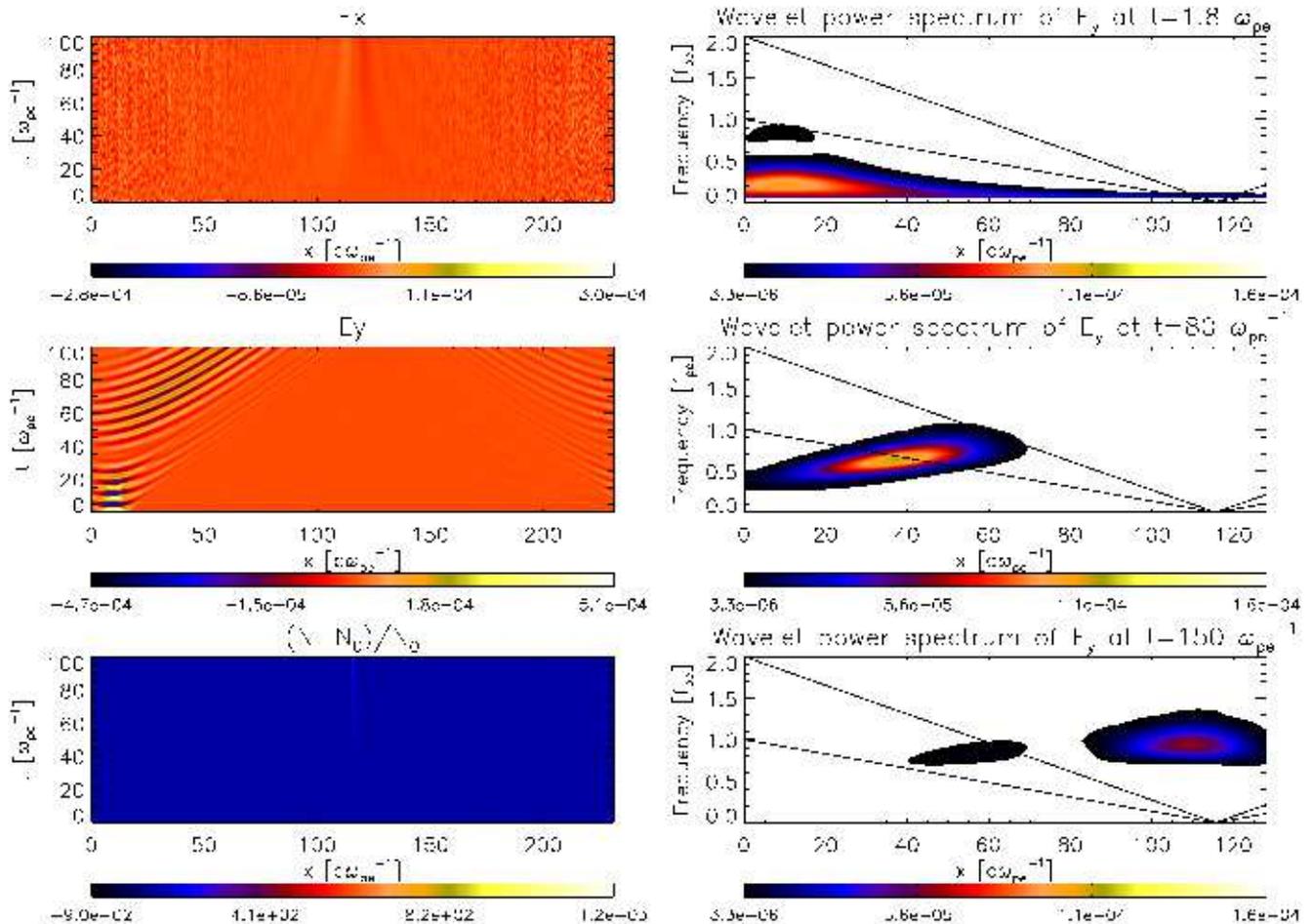}
\caption{\label{fig:tdds90} Left column: time-distance-plots of electric field components $E_x$ (top panel) and $E_y$ (middle panel), as well as change in density (bottom panel). Right: wavelet transform of $E_y$ for $t=1.8\omega_{pe}^{-1}$ (top panel), $t=80\omega_{pe}^{-1}$ (middle panel), $t=150\omega_{pe}^{-1}$ (bottom panel). Note that the background for the wavelet transform was set to white colour and does not refer to maximum amount of emission on the sides of the plots. Further, the black lines track the local plasma frequency and its second harmonic}
\end{figure*}
%----------------------------------------------------------------------------
\begin{figure}[htbp]
\includegraphics[scale=0.49]{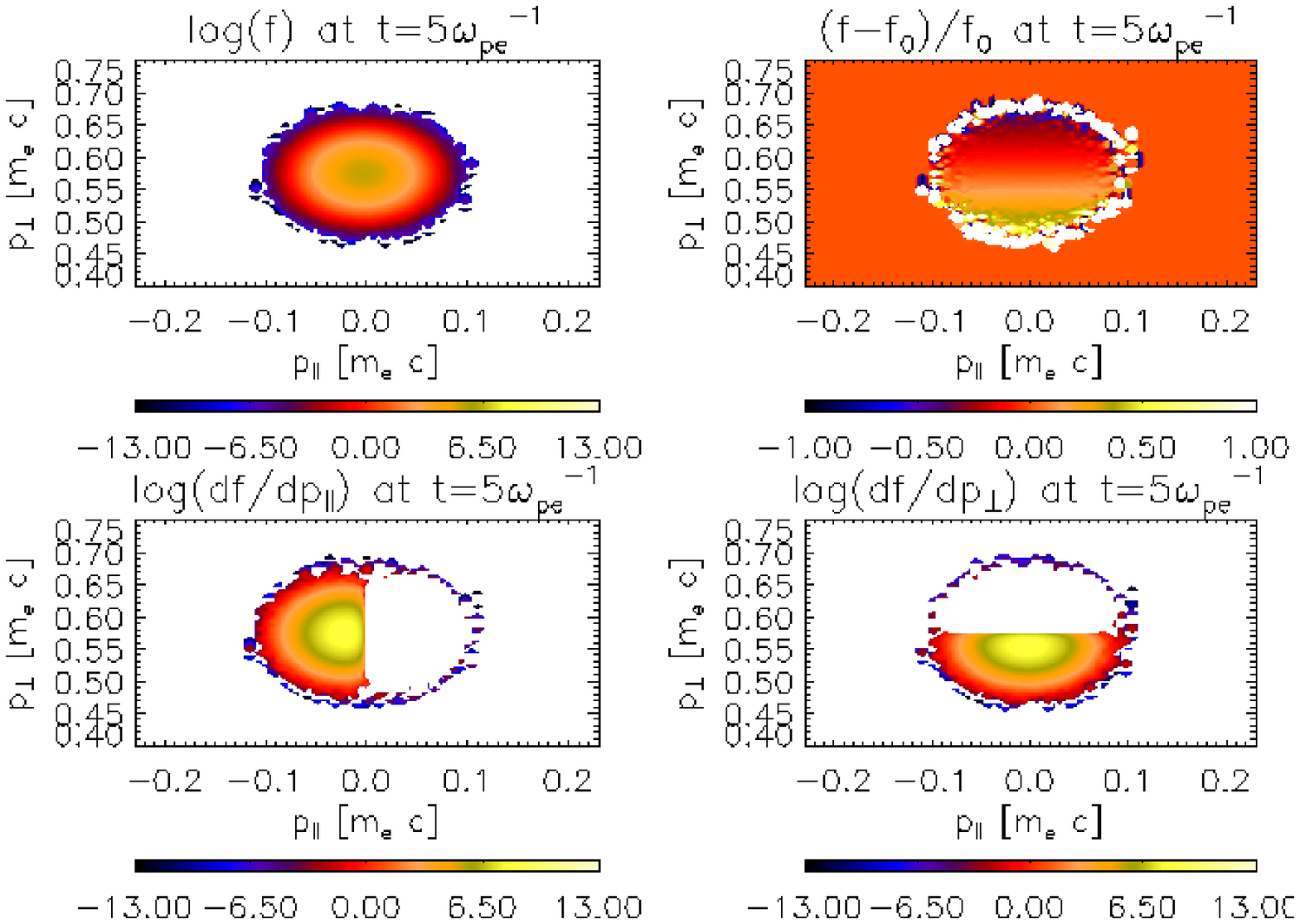}
\caption{\label{fig:dfpp} top left: $f(p_{\parallel},p_{\perp},t)$. top right: $\frac{f(t)-f(0)}{f(0)+\epsilon}$. bottom left: $\frac{\partial f(t)}{\partial p_{\parallel}}$. bottom right: $\frac{\partial f(t)}{\partial p_{\perp}}$ at $t=5\omega_{pe}^{-1}$. Gradients are shown in logarithmic scales, which cannot show negative values, therefore, they appear white as the background. }
\end{figure}
%----------------------------------------------------------------------------
Fig.\ref{fig:dfpp} is a representative snapshot of the investigated quantities of the electron distribution function at $t=5\omega_{pe}^{-1}$. The time evolution of the wavelet transform of the transverse electric component $E_y$ is shown in movie 1 \cite{mov:ref} (including lines to indicate the local plasma frequency, its second harmonic and the z-mode cut-off at $\omega_{z}=-\frac{1}{2}\omega_{ce}+\frac{1}{2}\sqrt{\omega_{ce}^2+4 \omega_{pe}^2}$), while the respective evolution of the distribution function is given in movie 2 \cite{mov:ref}. We can see from the distribution function, that the requirement for cyclotron maser emission, $\frac{\partial f}{\partial p_{\perp}}$, is fulfilled at all times. This generates waves at the electron cyclotron frequency, which can be seen in the wavelet transforms. Those frequencies are too far below the plasma frequency and do not allow the wave to propagate. It, therefore, decays again. The quick generation and decay of waves at $0 \le t \lesssim 30\omega_{pe}^{-1}$ appears as a pulsation. The frequency of this pulsation is found to be roughly twice the relativistic cyclotron frequency. It is likely that an energy exchange between beam and field, as shown in Ref.\cite{2012ApJ...752...60Y}, is happening at this stage, as the distribution function shows subtle but complex dynamics in this time interval. Eventually, at $\approx 40\omega_{pe}^{-1}$, a stable wave packet is formed that rises in frequency, while it slowly moves away from the injection region. This is indicative of a mode coupling process on the density gradient. At some instance, $(\omega,k)$ of the emitted cyclotron emission coincides with that of the z-mode. An effect similar to the one discussed in e.g. Figs.1 and 2 of Ref.\cite{1996PhRvE..53.6028C}. When the stable wave packet is formed, the distribution function shifts towards momenta higher than its initial maximum at $\approx 0.5775 m_ec$ and remains rather stable. At $t \approx 70\omega_{pe}^{-1}$ the wave packet reaches the lower cut-off frequency of the z-mode, at this point the distribution function shifts to lower momenta again. The wave packet keeps increasing its frequency until it reaches roughly the plasma frequency at the beam injection point.\\
We calculate the EM field energy by use of the Poynting theorem, $w(t)=\int [ \epsilon_0 {\bf E}(x,t)^2/2 + {\bf B}(x,t)^2/2\mu_0 ]dx$. We take care to exclude contributions from the background magnetic field $B_0$. We relate the field energy to the initial kinetic energy of the beam, $E^{beam}_{kin}(0)=7.89 \times 10^{-3}$J. The overall efficiency is $\approx 10^{-3} E^{beam}_{kin}(0)$, as shown in Fig.\ref{fig:EmEk}.\\
We also find the wave to be of left-handed polarization \cite{PoP}. In the case of ($\omega_{ce}/\omega_{pe} \ll 1$), the cyclotron maser is known to generate z-mode waves, at frequencies that are harmonics of the cyclotron frequency \cite{1986ApJ...307..808W}. The z-mode is left-hand polarized for $\omega<\omega_{pe}$ \cite{2000PhPl....7.3167W}. In movie 1 \cite{mov:ref}, it is shown that the wave packet is stabilized just below the local plasma frequency, therefore, it is no surprise to find the emitted wave to be of left-hand polarization.\\
In Ref.\cite{2011PhPl...18e2903T}, it was shown that a non-gyrotropic beam injection into plasma can generate EM emission. Ref.\cite{PoP} has shown that only the perpendicular momentum component of the injected electron beam causes the generation of EM emission. The main focus of Ref.\cite{PoP} was a parametric study, i.e. how different beam injection pitch angles and different gradients affect the EM emission generation. Here the main goal is to focus on the physics of the EM emission \textit{generation mechanism}. Thus, we carried out a 1.5D PIC simulation of a beam of relativistic, hot electrons injected with only a momentum component perpendicular to the background magnetic field into a maxwellian plasma with a density gradient. The choice of 1.5D turns off all EM emission generated by the classic plasma emission mechanism for all pitch angles, as it needs at least 2 spatial dimensions to work. The lack of parallel beam momentum meant that there was no electrostatic Langmuir wave generation, yet there was still EM emission due to the cyclotron maser. Therefore, there is no need for generation of Langmuir waves in order to generate EM emission via non-gyrotropic beam injection. The cyclotron maser generates emission at the electron cyclotron frequency, which - in the beam injection region of our simulation - is far below the plasma frequency, hence cannot propagate and decays. Generation and decay result in a pulsating emission generation until a stable wave packet is formed, that mode couples on the density gradient to what is likely to be a z-mode, until it reaches cut-off frequencies that allow for propagation. It may well be, that the quasi-linear relaxation, which would result in a plateau forming of the distribution function and eventual shutting down of the emission mechanism, happens on the same time scales as the inverse growth rate. Calculations of growth rates are not part of this study. We do not see a continuous generation of emission, because the injected beam has no temporal extent (as it would have in a real solar type III burst situation) and beam electrons are not replenished, i.e. injection happens only at $t=0$. Further, a propagating beam would trigger the presented mechanism along its trajectory and, hence, produce the characteristic type III burst shape in the dynamical spectrum. Naturally, real life is not 1.5D, and generally there will %---------------------------------------------------------------------------
%\begin{figure}[h]
%\includegraphics[scale=0.48]{EmofEkPPL.eps}
%\caption{\label{fig:EmEk} Total EM field energy, $w(t)$, normalized to the initial beam kinetic energy, $E^{beam}_{kin}(0)$. }
%\end{figure}
%---------------------------------------------------------------------------
be contributions from classic plasma emission mechanisms and the antenna radiation (if density cavities are formed) as well as the cyclotron maser.
However, in cases of strong magnetic fields and straight field lines, situations can be well approximated by a 1.5D model. %---------------------------------------------------------------------------
\begin{figure}[h]
\includegraphics[scale=0.48]{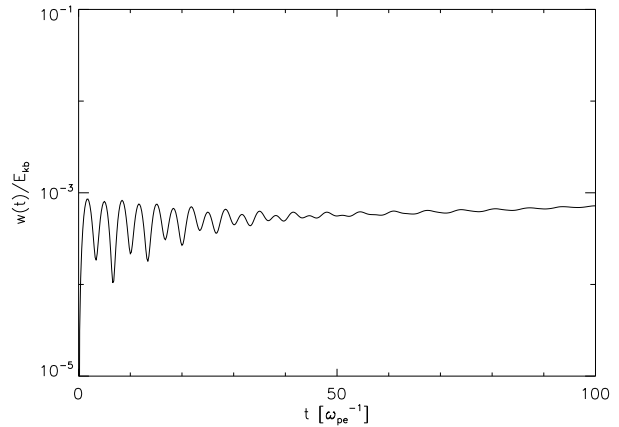}
\caption{\label{fig:EmEk} Total EM field energy, $w(t)$, normalized to the initial beam kinetic energy, $E^{beam}_{kin}(0)$. }
\end{figure}
%---------------------------------------------------------------------------
In any case, there is likely a perpendicular beam momentum component, that will contribute to the overall emission. The interplay between EM emission generation mechanisms needs to be studied for various parameter spaces, but is not part of this work. Since the proposed mechanism has many applications (e.g. magnetrons, radar), we believe it could be interesting for the physics community at large.\\
The authors are financially supported by the HEFCE-funded SEPNET. D.T.'s research is supported by The Leverhulme Trust Grant RPG-311 and STFC Grants ST/J001546/1 and ST/H008799/1.\\
{\bf note added in proofs:} After this work was complete, we became aware of the following: (i) when a case of $T_{background}=3\times10^8$ is studied, eliminating the electron cyclotron maser (ECM) instability as $\frac{\partial f}{\partial p_{\perp}}<0$, $\forall {\bf p}$, EM emission identical to the one in this study still occurs; (ii) when a ring-shaped (in $p_y$ and $p_z$) beam distribution is considered, implying ${\bf j_{\perp}}=0$, no EM emission is generated. These findings indicate that both effects - the ECM instability and EM emission from transverse currents - are present (and competing). It is probable that the ECM instability growth rate is so small that it cannot develop by the end of the simulation. Calculation of the growth rates commensurate to physical parameters of type III bursts and the ring distribution will be published elsewhere.
%\bibliography{PPL}% Produces the bibliography via BibTeX.

\end{document}